# Accelerated design of linear-superelastic Ti-Nb nanocomposite alloys with ultralow modulus via high-throughput phase-field simulations and machine learning

Yuquan Zhu[1], Tao Xu[1,2*], Qinghua Wei[1], Hongxin Yang[2], Takahiro Shimada[3], Takayuki Kitamura[3] and Tong-Yi Zhang[1**]

[1]*Materials Genome Institute, Shanghai University, Shanghai 200444, China*

[2]*Ningbo Institute of Materials Technology and Engineering, Chinese Academy of Sciences, Ningbo 315201, China*

[3]*Department of Mechanical Engineering and Science, Kyoto University, Nishikyo-ku, Kyoto 615-8540, Japan*

**Abstract**

The controlled design of martensitic transformation (MT) to achieve specific properties is crucial for the innovative application of shape memory alloys (SMAs) in advanced technologies. Herein, we explore and design the MT behaviors and the mechanical properties of Ti-Nb nanocomposites by combining high-throughput phase-field simulations and machine learning (ML) approaches. Based on the systematic phase-field simulations, we obtain data sets of the mechanical properties for various nanocomposites constructed by four macroscopic degrees of freedom, which can be employed to design and optimize the microstructures for different applications. To accelerate the phase-field screening of the desired metallic biomaterials, a ML assisted strategy is adopted to perform multi-objective optimization of the mechanical properties, through which promising nanocomposite configurations are pre-screened for the next set of phase-field simulations. With the ML guided simulations, an optimized candidate composed of Nb-rich matrix and Nb-lean nanofillers that exhibits a combination of unprecedented mechanical properties, including ultra-low modulus, linear super-elasticity, and near-hysteresis-free is designed. The exceptional mechanical properties in the nanocomposite originate from optimized continuous MT rather than a sharp first-order transition, which is common in typical SMAs. This work provides a new computational approach and design concept for developing novel functional materials with extraordinary properties.

## 1. Introduction

Titanium-based shape memory alloys (SMAs), such as Ti-Nb alloys, are an important class of smart materials that possess shape memory effect (SME) and pseudoelasticity (PE) [1], as well as high specific strength, excellent corrosion resistance, superior biocompatibility [2-3], etc. These fascinating merits are widely exploited for industrial [4] and biomedicine applications [5]. In general, the SME and PE originate from temperature- or stress-induced reversible martensitic transformation (MT) [6]. Being a strong first-order structural transition [7], MT usually suffer from a large stress-strain hysteresis that consists of an initial true elasticity stage with a high Young's modulus (> 80 GPa) and a following stress-plateau that is associated with reversible structural transition. As a result, low efficiency and poor position control of the SMA actuators are generally observed due to the large hysteresis and strong non-linearity [8]. Moreover, as potential metallic biomaterials, a low Young's modulus comparable with those of natural human bones (~ 20 GPa [9]) is essential for Ti-Nb alloys to avoid the "stress shielding effect" [10] and the resulting bone degradation. Although several efforts have been recently made to optimize the mechanical responses of metallic alloys, e.g., modulation of the components and concentrations [11-12], defect engineering [13-14], introduction of elastic-inelastic strain matching [15-16], and grain refinement [17-18], the deliberate design and control of MTs for a combination of specific properties such as low modulus, linear super-elasticity, and free-hysteresis is still highly desired for various advanced biomedical and engineering applications.

From the perspective of the Landau thermodynamic theory, many physical quantities of ferroic materials are associated with free-energy derivatives with respect to certain thermodynamic variables. For instance, the dielectric permittivity and piezoelectric coefficient of ferroelectrics are determined by second derivatives of the Gibbs free energy density with respect to polarization [19-20], i.e., curvature of the thermodynamic energy profile. Based on this, ultrahigh piezoelectricity [20] and ultrahigh energy density dielectrics with minimized hysteresis [21] have been designed in ferroelectrics via flattening the thermodynamic energy landscape, which is achieved by judiciously introducing nanoscale structural heterogeneity or nanodomains. Such structure manipulations can be realized by the addition of dopants or the fabrication of solid solutions, as have been experimentally demonstrated [22-23]. In analogy to dielectric properties in ferroelectrics, the mechanic properties (e.g., elastic modulus) of ferroelastic metals such as SMAs are also closely related to the second derivatives of thermodynamic energy [24-26]. This suggests the possibility of optimizing the overall mechanical performance of SMAs as well by nanoscale structure manipulation, including the rational nanocomposite design in Ti-Nb alloys.

Conventionally, time-consuming experimental searches by the trial-and-error process are employed to find new materials for specific applications. Computational methods can speed up these investigations and are now playing an increasingly important role in the search for new materials or structures with tailored properties and novel functionalities. The emerging approach of high-throughput (HTP) material design has been recognized as a powerful tool in this field without requiring initial experimental synthesis and characterization [27]. There have been many inspirational HTP first-principles works, from which direct links between atomic-scale information and macroscopic functionalities are established. These approaches have shown promise in the discovery of new piezoelectric and dielectric materials [28], thermoelectric materials [29], magnetic material[30], and so on. On the other hand, the origin of material properties resides in not only chemical constituent itself but also mesoscale morphological and microstructure evolutions [31]. However, mesoscale phase-field simulations in a high throughput manner that allow for the investigation of microstructure effects are rather limited [32-34]. The development of HTP mesoscale calculations is not only indispensable for the accelerated characterization of microstructure evolutions but also a boost for data- and modeling-driven discovery of new materials and structures. In this work, we develop a highly efficient phase-field simulation framework to optimize the mechanical response of Ti-Nb SMA nanocomposites without introducing additional elements and seek the optimal microstructure with desired properties. Based on the combination of HTP phase-field simulations and machine learning (ML), a nanocomposite with a perfect combination of ultralow modulus, quasi-linear elasticity, and near-zero hysteresis is screened out, which shows great potential for biomedical material applications.

*2.1 Simulation method*

The Ti-Nb-based alloys undergo diffusionless and reversible MT between the cubic $\beta$-austenite (point group $m\bar{3}m$) and orthorhombic $\alpha''$-martensite (point group $mmm$) phases [35] during the thermomechanical loading, as shown in Fig. 1a. This cubic-to-orthorhombic MT is associated with six variants [36-37]. For simplicity, we construct a two-dimensional nanocomposite phase-field model within the plane stress condition. As a result, there are four possible variants (V1–V4) corresponded to four different stress-free transformation strain (SFTS) tensors after dropping the out-of-plane strain. All the SFTSs can be found in the Supporting Information. In the phase-field model, a continuous structure order parameter $\eta_i$ ($i$ = 1 - 4) is employed to characterize the MT, with $\eta_i = 0$ and $\eta_i = \pm 1$ representing austenitic and $i$-th martensitic variant, respectively. The total free energy of the system $F$ consists of the Landau free energy $f_{ch}$, the gradient energy $f_{gr}$, and the elastic energy $f_{el}$ [38]:

$$F = \int (f_{ch} + f_{gr} + f_{el})\, dV. \qquad (1)$$

The microstructure evolution of the MT is governed by the time-dependent Ginzburg-Landau (TDGL) equation [39]:

$$\frac{\partial \eta_i(r,t)}{\partial t} = -L \frac{\delta F}{\delta \eta_i(r,t)} + \zeta_i(\boldsymbol{r},t), i = 1\sim 4, \qquad (2)$$

where $L$ is the kinetic parameter, and $\zeta_i(\boldsymbol{r},t)$ is the Stochastic-Langevin noise term accounting for the effect of thermal fluctuations [40]. Full details of the simulation methods and the expressions for different energies are provided in the Supporting Information.

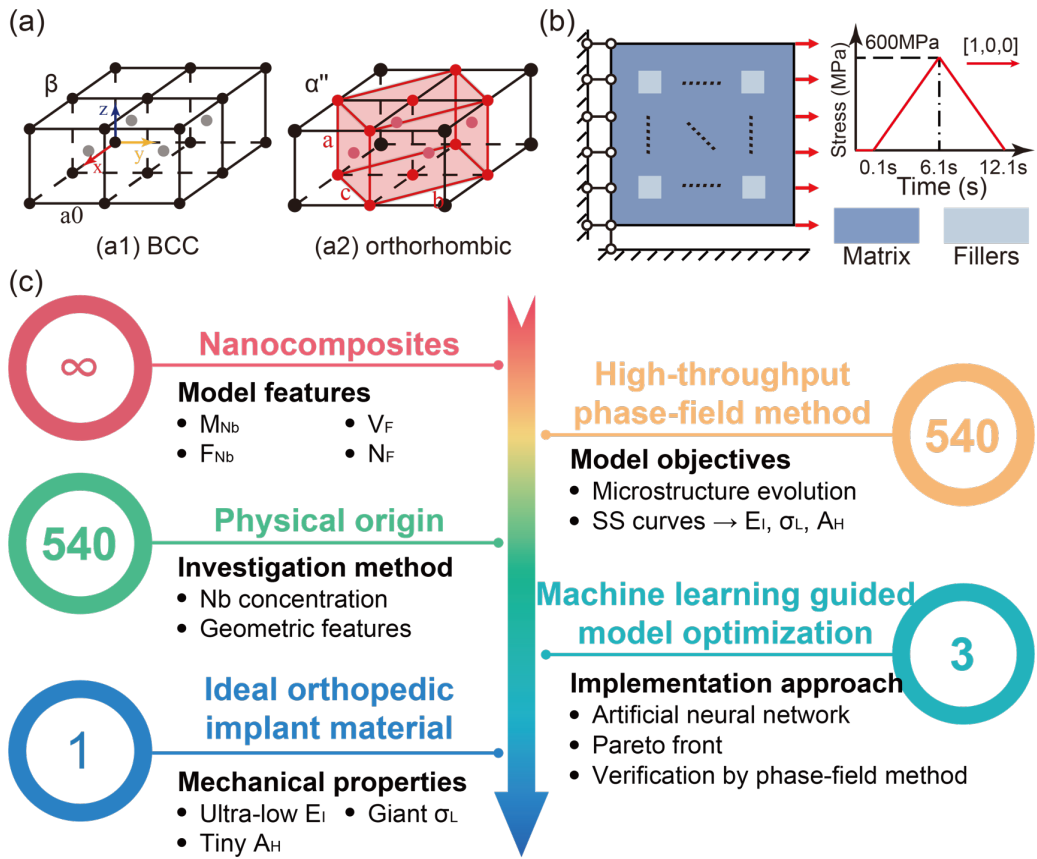

**Fig. 1.** (a) Crystal structure of parent (BCC) and martensitic (orthorhombic) phase. (b) Distribution of Nb-lean nanofillers in the square-array Ti-Nb nanocomposite and loading-unloading process. (c) HTP phase-field simulation framework for the Ti-Nb nanocomposite.

*2.2 Model and procedure*

We design a series of Ti-Nb square-array nanocomposite structures [41] consisting of different Nb-lean nanofillers that are uniformly and coherently embedded in a Nb-rich Ti-Nb alloy matrix (Fig. 1b). The motivation for the choice of such a nanostructure is the advancements in fabrication techniques of composites and their various associated phenomena [42-44]. The Ti-Nb nanocomposite has four structural feature variables (macroscopic degrees of freedom) related to the macroscopic mechanical properties, namely the Nb concentration ($C_{Nb}$) of the matrix ($M_{Nb}$) and nanofillers ($F_{Nb}$), volume fraction of nanofillers ($V_F$), and number of nanofillers ($N_F$). $M_{Nb}$ and $F_{Nb}$ have concentration ranges of 15%–20% and 5%–10%, respectively, with a 1% interval. For modeling purposes, we choose five groups of $V_F$ varying from 1.56% to 39.06%, which correspond to the total area of nanofillers $8^2$ nm$^2$, $16^2$ nm$^2$, $24^2$ nm$^2$, $32^2$ nm$^2$, and $40^2$ nm$^2$. Correspondingly, three groups of $N_F$ are defined: 4, 16, and 64. The computational framework for the design of the Ti-Nb nanocomposite is represented schematically in Fig. 1c. We start by specifying the above four features ($M_{Nb}$, $F_{Nb}$, $V_F$, and $N_F$) with automatic nanocomposite modeling, which yields 540 candidates from a combinatorial point of view. Although the nanofillers could be randomly distributed in the matrix, we restrict the nanofillers to a square nanocomposite structure [41], a typical Archimedean lattice structure, to reduce the computational complexity. We then carry out high efficiency phase-field simulations for the established nanocomposite models and compute their microstructure evolutions and stress-strain (SS) curves under mechanical stress along the [100] direction as shown in Fig. 1b. The quantitative analyses of the SS curves are subsequently conducted, and the main mechanical properties are extracted, including the apparent incipient Young's modulus ($E_I$), the elastic stress limit ($\sigma_L$), and the hysteresis area ($A_H$). Based on the data sets from HTP phase-field simulations, we employ ML strategy to perform multi-objective optimization of the mechanical properties for the nanocomposites. The recommended candidates are further verified by the phase-field simulations and the underlying mechanisms are also clarified.

*3.1. Overview of high-throughput phase-field results*

Based on the HTP phase-field simulations, we plot the obtained four features dependence of the mechanical properties of the nanocomposites in Fig. 2, in which $V_F$ increases from 1.56% to 39.06% on each row and $N_F$ decreases from 64 to 4 on each column. It can be seen that $E_I$ and $\varepsilon_L$ of the nanocomposites are widely distributed within 10–35 GPa and 0%–3%, respectively. This tunable and variable modulus indicates that engineering the chemical composition and geometry of the nanocomposite offers great flexibility to turn the mechanical properties of the materials. According to the map shown in Fig. 2a, there is a rough trend that nanocomposites with larger $V_F$ exhibit lower $E_I$, although the Nb-lean nanofillers is assumed to have the same elastic constants as that of the matrix. It is also interesting that $F_{Nb}$ dramatically reduces $E_I$ at a lower $F_{Nb}$ ($F_{Nb}$ < 8%) (see panels a3–a5, a8–a10, and a13–a15) and higher $V_F$ ($V_F$ > 14.06%), while $M_{Nb}$ and $N_F$ are negligible on $E_I$ for the whole $V_F$ range. These results indicate that coupling between nanofillers and matrix with different $C_{Nb}$ gives rise to $E_I$ superior to their inherent ones. As displayed in Fig. 2b, $\sigma_L$ also decreases with increasing $V_F$, which is consistent with the conclusion of $E_I$, indicating that the embedding of nanofillers changes the mechanical properties drastically. In comparison, $\sigma_L$ increases with increasing $M_{Nb}$, indicating that the Nb-rich matrix effectively maintain the linear elasticity of the nanocomposites. The opposite variation tendencies are observed for $E_I$ and $\sigma_L$ with $F_{Nb}$ and $M_{Nb}$, indicating that $F_{Nb}$ and $M_{Nb}$ play different roles in regulating the mechanical properties of the nanocomposites. Moreover, the calculated $A_H$ of the nanocomposites spans a large range from 0 MJ/m$^3$ to 3 MJ/m$^3$ for the different features (Fig. 2c), indicating that the embedding of nanofillers greatly reduces the energy loss during the loading-unloading cycle. Although the regulated effect of $N_F$ is not as obvious as other features, it can be seen from Fig. 2a to 2c that $N_F$ is of great significance in fine-tuning mechanical properties of nanocomposites. All these results indicate that the mechanical properties of nanocomposites are nonlinearly affected by all four features.

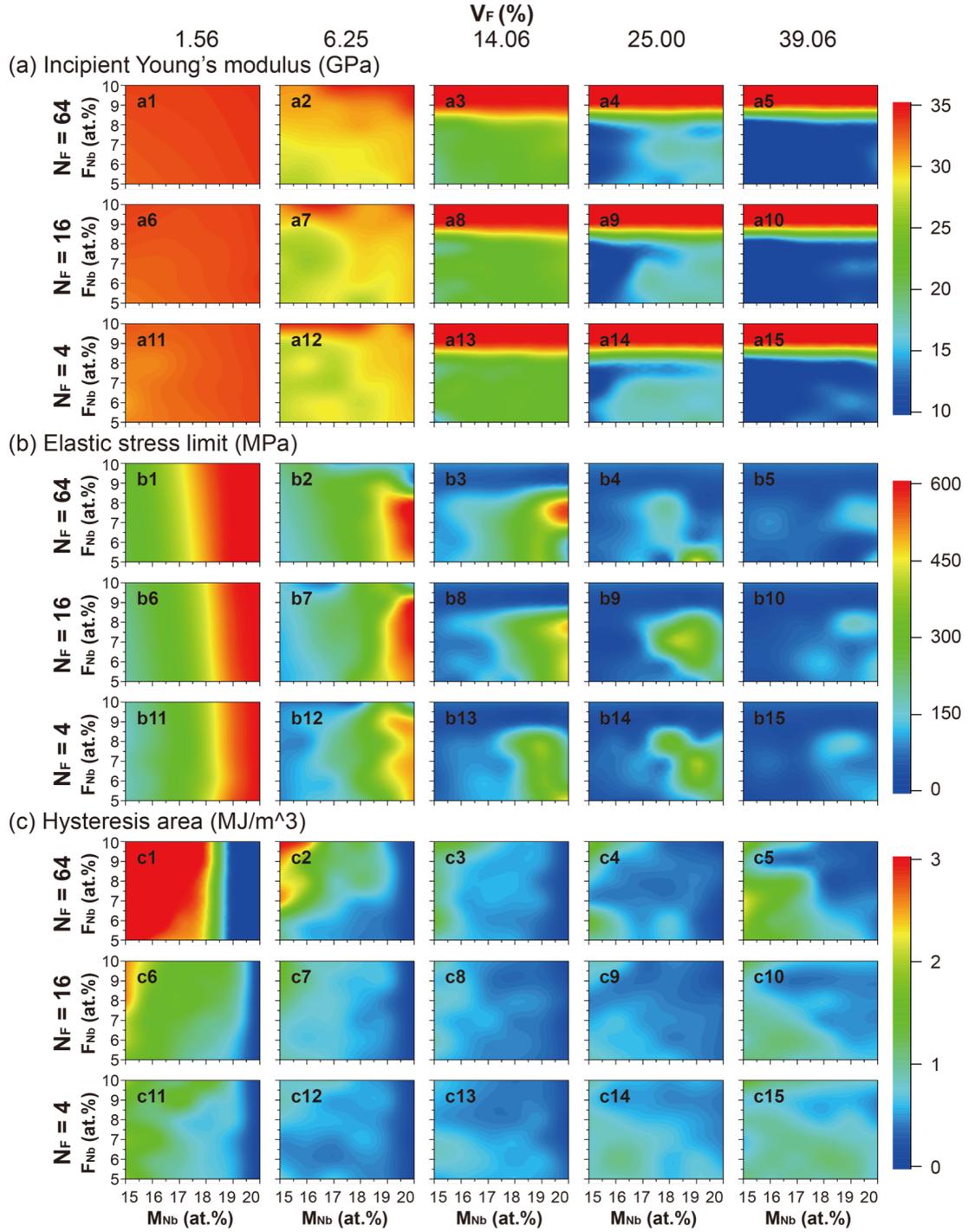

**Fig. 2.** Mapping results of the HTP phase-field simulations of (a) the apparent incipient Young's modulus, (b) the elastic stress limit, and (c) the hysteresis area.

### 3.2. Effect of the coherent nanocomposite

The widely tunable and variable mechanical responses of the nanocomposites could be attributed to the change in the MT nucleation critical stress and the local stress fields associated with the inserted
9

nanofillers. Fig. 3a1 plots the Landau free energy density curves as a function of the order parameter for the Ti-Nb alloys with different $C_{Nb}$ at 300 K, in which the martensitic phase gradually transformed from an unstable state to a stable state with decreasing $C_{Nb}$. It is evident that the martensitic phase is stable in the Nb-lean cases ($C_{Nb} < 8\%$), while in the Nb-rich cases ($C_{Nb} \geqslant 8\%$), the martensitic phase could not nucleate without the aid of external loads. The critical stress for the stress-induced martensitic nucleation versus $C_{Nb}$ is further calculated (see Fig. 3a2) and the results demonstrate that the critical stress increases exponentially with increasing $C_{Nb}$. For the heterogeneous nanocomposites, martensite thus preferentially nucleates in Nb-lean nanofillers and generates local stress fields associated with the $C_{Nb}$-dependent SFTS tensors (see Fig. 3a2) to promote the growth of martensite into matrix, while the high critical stress in the Nb-rich matrix will confine the extent of growth. To explore the ability that Nb-rich matrix inhibits the spread of martensite from nanofillers into matrix, we perform phase-field simulations for the simplified model as schematically shown on the left side of Fig. 3a3. It consists of 8nm$^2$ martensite particle with Nb concentration of $f_{Nb}$ (denoted as the green square) and the adjacent 8nm$^2$ austenitic matrix with Nb concentration of $m_{Nb}$ (denoted as the orange square), corresponding to martensite volume fraction of 0.5. After relaxation, the interface between martensite and austenite moves toward the particle or matrix depending on $f_{Nb}$ and $m_{Nb}$. Different combinations of $f_{Nb}$ and $m_{Nb}$ will generate various martensite volume fractions as shown in Fig. 3a3. The heatmap of martensite volume fraction after relaxation is divided by a dashed line into suppressed and non-suppressed region, which is highlighted in blue and red colors, respectively. These results thus indicate that $F_{Nb}$ ($f_{Nb}$) and $M_{Nb}$ ($m_{Nb}$) significantly mediate the Landau free energy and SFTS, and further regulate the MT nucleation critical stress and spread of martensite in the nanocomposites.

Furthermore, the geometric features (i.e., $N_F$ and $V_F$) modulate the strength and distribution of the generated stress field. We perform simplified simulations to investigate the local stress field $\sigma_{11}$ (Fig. 3b) generated by the lattice mismatch between the embedded martensitic fillers (V3 with $F_{Nb}$ = 7%) and the matrix with different $N_F$ and $V_F$, and the results are summarized in Fig. 3c. We observe that an increase in $V_F$ resulted in the gradual expansion of the local stress distribution range, such that $V_F$ greater than 25% produces a second stress distribution peak around 400 MPa. In addition, the increase of $V_F$ not only increases the maximum value of the local stress field (represented by the red dashed line) but also significantly increases the probability that the local stress is greater than the critical stress of martensite spread into matrix, as indicated by the blue dashed line in Fig. 3c for the median value (i.e., 600 MPa). Moreover, $N_F$ also regulates the distribution and probability density of the local stress field. Contrary to $V_F$, a smaller $N_F$ increases the maximum value of the local stress field and the probability of $\sigma_{11} \geqslant$ 600 MPa. Hence, $V_F$ and $N_F$ effectively affect the magnitude and extent of the local stress field. Therefore, all four features regulate the nucleation and spread of MT process by altering the MT nucleation critical stress and the local stress field distribution.

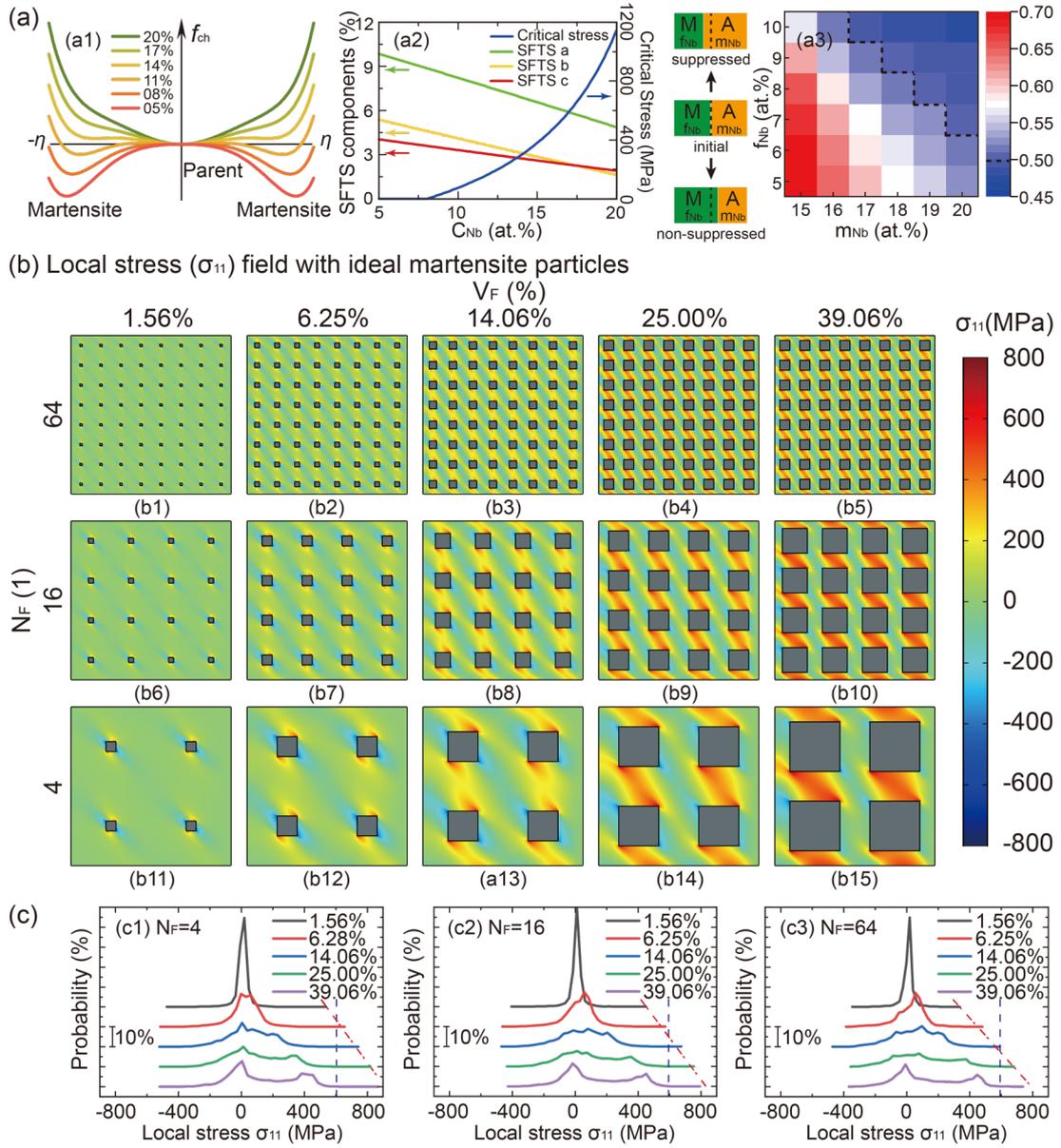

**Fig. 3.** Schematic diagrams for understanding the physical origin of fillers to achieve the specified properties. (a1) Landau free energy curves with different $C_{Nb}$, (a2) SFTS tensors component, and MT critical stress with different $C_{Nb}$, and (a3) the ability that Nb-rich matrix inhibits the spread of martensite in fillers into matrix determined by $F_{Nb}$ and $M_{Nb}$. (b) Local stress field on the [100] direction of Ti-Nb nanocomposites of different $V_F$ and different $N_F$ and (c) local stress field probability distribution with $V_F$ = 1.56%, 6.28%, 14.06%, 25.00%, and 39.06% for different $N_F$.

### 3.3. Machine learning guided optimization of the nanocomposites

As previously discussed, reducing the Young's modulus while maintaining relatively large elastic strain limit ($\varepsilon_L$) is critical for ideal orthopedic implant. Through the brute-force phase-filed simulations of the initial hundreds of nanocomposites, we have obtained the structures with widely tunable mechanical properties. However, an exhaustive search for the optimal one is intractable and computationally expensive. To remedy this, we apply the artificial neural network (ANN) method using the WEKA platform [45] to achieve the highly efficient multi-objective optimization (i.e. $E_I$ and $\sigma_L$) of the nanocomposites. On the basis of the data sets from HTP phase-field simulations, we train ANN model to predict the mechanical properties of unexplored nanocomposites with refined structural feature variables. In order to avoid samples with poor performance dominating the construction of the ML model, 198 samples with excellent performance (here we select the condition of $\sigma_L \geqslant 20 \times E_I - 200$) in the HTP phase-field simulations are selected as training data for the ANN model (see Fig. 4b). We estimate the generalization ability of the models and optimize the hyper-parameters of ANN utilizing the 10-fold Cross Validation method. The criterion for prediction accuracy is measured by the correlation coefficient (R), which are defined by

$$R = \frac{\Sigma(y_i - \bar{y})(\hat{y}_i - \bar{\hat{y}})}{\sqrt{\Sigma(y_i - \bar{y})^2 \Sigma(\hat{y}_i - \bar{\hat{y}})^2}} \qquad (3)$$

where $y$, $\bar{y}$, and $\hat{y}$ represent the actual, average, and predicted values, respectively. On the basis of ANN predictions, we could obtain the competing mechanical properties, $E_I$ and $\sigma_L$, of a set of nanocomposites in the search space. To determine the nanocomposites with a desirable combination of the two properties for next simulations, a multi-objective design strategy based on the Pareto front [46] is employed, as shown in Fig. 4a. Two targets, property 1 and property 2, are set in two-dimensional space, in which the red star can be considered as an ideal performance. The Pareto front defines a characteristic boundary consisted of a set of data points which are not dominated by any other points in the data set. In the data sets with two properties ($P_1$, $P_2$), if $\forall$ k ∈ {1,2}: $P_{ki} \leq P_{kj} \lor \exists$ k ∈ {1,2}: $P_{ki} < P_{kj}$, then the point $P_i$ is said to be dominated by point $P_j$ ($P_{j1}$, $P_{j2}$). On the Pareto front, we use a selection strategy to find the three "best" configurations as the candidate and measure them by the phase-field simulation. Our selection strategy transforms the two-objective optimization problem into a single-objective optimization problem by calculating the normalized distance δ from a point in virtual space to the target ($E_I^t$ = 10GPa, $\sigma_L^t$ = 500MPa), and the normalized Euclidean distance δ is given by,

$$\delta = \sqrt{\left(\frac{E_I - E_I^t}{35}\right)^2 + \left(\frac{\sigma_L - \sigma_L^t}{600}\right)^2} = \sqrt{\left(\frac{E_I - 10}{35}\right)^2 + \left(\frac{\sigma_L - 500}{600}\right)^2} \tag{4}$$

Fig. 4c1-c2 show the cross-validation performance of $E_I$ and $S_L$ obtained by ANN algorithm. As we can see, the ANN model has a great cross-validation performance for $E_I$ (R = 0.9947) and $\sigma_L$ (R = 0.9823). In order to predict new configurations with better properties, we created a total of 3794 virtual samples around the training data points by finetuning the $M_{Nb}$ and $F_{Nb}$ with a step size of $\Delta C_{Nb}$=0.2%. According to the hyperparameters determined by 10-fold cross-validation, we use all the training data to obtain two ANN models to predict the $E_I$ and $\sigma_L$ of the virtual space. The virtual space based on the model predictions is visualized in Fig. 4d, along with the training data points and Pareto front of the virtual space. According to our selection strategy shown in Fig. 4a, three new configurations with improved properties are screened out in the virtual space, which are selected as promising nanocomposite configurations for the next set of phase-field simulations. Their predicted and calculated mechanical properties are shown in Table 1. The predicted $E_I$ is lower than the calculated value with an average prediction error of 2.7%, while the predicted $\sigma_L$ is higher than the calculated value with an average prediction error of 0.6%, indicating that the ML model yield accurate predictions for the unexplored nanocomposites. The calculated properties of the three recommended configurations (denoted as red pentagons) are compared with those from HTP phase-field simulations (denoted as blue dots) in Fig. 4e. A clear improvement can be observed for the recommended configurations and the dark red pentagon sample with the most obvious performance improvement is selected as the configuration of ideal orthopedic implant. Compared with the result of HTP phase-field

simulations, the normalized Euclidean distance δ between the optimized configuration and the target performance is reduced from 0.230 to 0.204, decreasing by 11.3%. Thus, we could locate the configuration with the best compromise of lower $E_I$ and larger $\sigma_L$ with the aid of ANN. The optimized configuration is composed of Nb-rich TiNb matrix ($M_{Nb}$ = 18. 2%) and 25% volume fraction of Nb-lean TiNb square fillers ($F_{Nb}$ = 6.8%) with $N_F$ = 16, which will be discussed in detail below.

Table 1. Predicted and calculated value of properties of new nanocomposite

| $N_F$ (1) | $V_F$ (%) | $F_{Nb}$ (%) | $M_{Nb}$ (%) | Predicted | Calculated | Prediction Error |
|---|---|---|---|---|---|---|
| 16 | 25 | 18.2 | 6.8 | $E_I$ = 14.96GPa | $E_I$ = 15.42GPa | $e_E$ = 3.0% |
|  |  |  |  | $\sigma_L$ = 420.69MPa | $\sigma_L$ = 420.63MPa | $e_S$ = 0.0% |
| 16 | 25 | 18.2 | 6.6 | $E_I$ = 14.84GPa | $E_I$ = 15.26GPa | $e_E$ = 3.0% |
|  |  |  |  | $\sigma_L$ = 413.91MPa | $\sigma_L$ = 407.40MPa | $e_S$ = -1.6% |
| 16 | 25 | 18.2 | 7.2 | $E_I$ = 15.10GPa | $E_I$ = 15.44GPa | $e_E$ = 2.2% |
|  |  |  |  | $\sigma_L$ = 420.24MPa | $\sigma_L$ = 419.65MPa | $e_S$ = -0.1% |

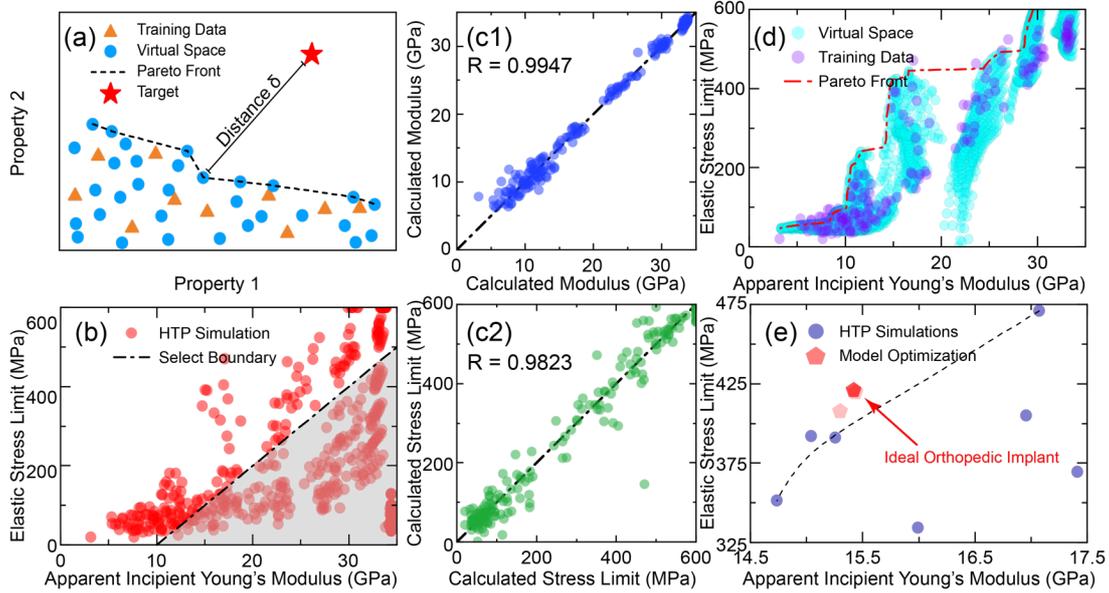

**Fig. 4.** (a) A schematic of Pareto front with two objectives. Our design strategy for selecting the best solution calculates the normalized distance δ from a point in the virtual space to the target. (b) The ML training data is selected according to the criteria of $\sigma_L \geq 20 \times E_I$ - 200. (c) Scattering plots of cross-validation performance presented by comparing the results of the ML and the training data. (c1) $E_I$; (c2) $\sigma_L$. (d) The Pareto front of two objectives and the comparison between Pareto front and 3794 virtual space which were expand from the selected training data. (e) The distribution of our new calculated points and training data derived from HTP phase-field simulations.

*3.4. Ideal orthopedic implant design strategy*

The comparison of $E_I$ and $\varepsilon_L$ of the constructed nanocomposites designed from the ML aided simulation with those of Ti2448 (Ti-24Nb-4Zr-8Sn in wt.%) [47], a typical metallic biomaterial, and other state-of-the-art systems proposed in the literature [48], are shown in Fig. 5a. It is readily seen that the candidate nanocomposite for the ideal orthopedic implant exhibits an ultralow $E_I$ relative to other reported systems, which is essential for metallic biomaterials. The nanocomposite also outperforms Ti2448 in terms of the relatively large elastic strain limit and small hysteresis area. The SS curve of the designed nanocomposite during the loading-unloading cycle together with that of uniform bulk Ti2448 is shown in Fig. 5c with the red curve and gray curve, respectively. The SS curve of Ti2448 features a typical characteristic of a large stress hysteresis with an obvious stress plateau, which is consistent with the experimental observations. By contrast, the stress plateau completely disappears in the SS curve of ideal orthopedic implant, and exhibits a quasi-linear elasticity and almost hysteresis-free mechanical behavior. Moreover, as compared to Ti2448, a dramatically reduced $E_I$ of 15.4 GPa, even lower than that of the nature bones (~20 GPa [9]), is achieved without sacrificing $\sigma_L$ (i.e., 420 MPa) in ideal orthopedic implant. Thus, the designed ideal orthopedic implant exhibits a perfect combination of linear superelasticity and ultralow $E_I$ while maintaining relative high strength, which are superior to those of the reported advanced systems.

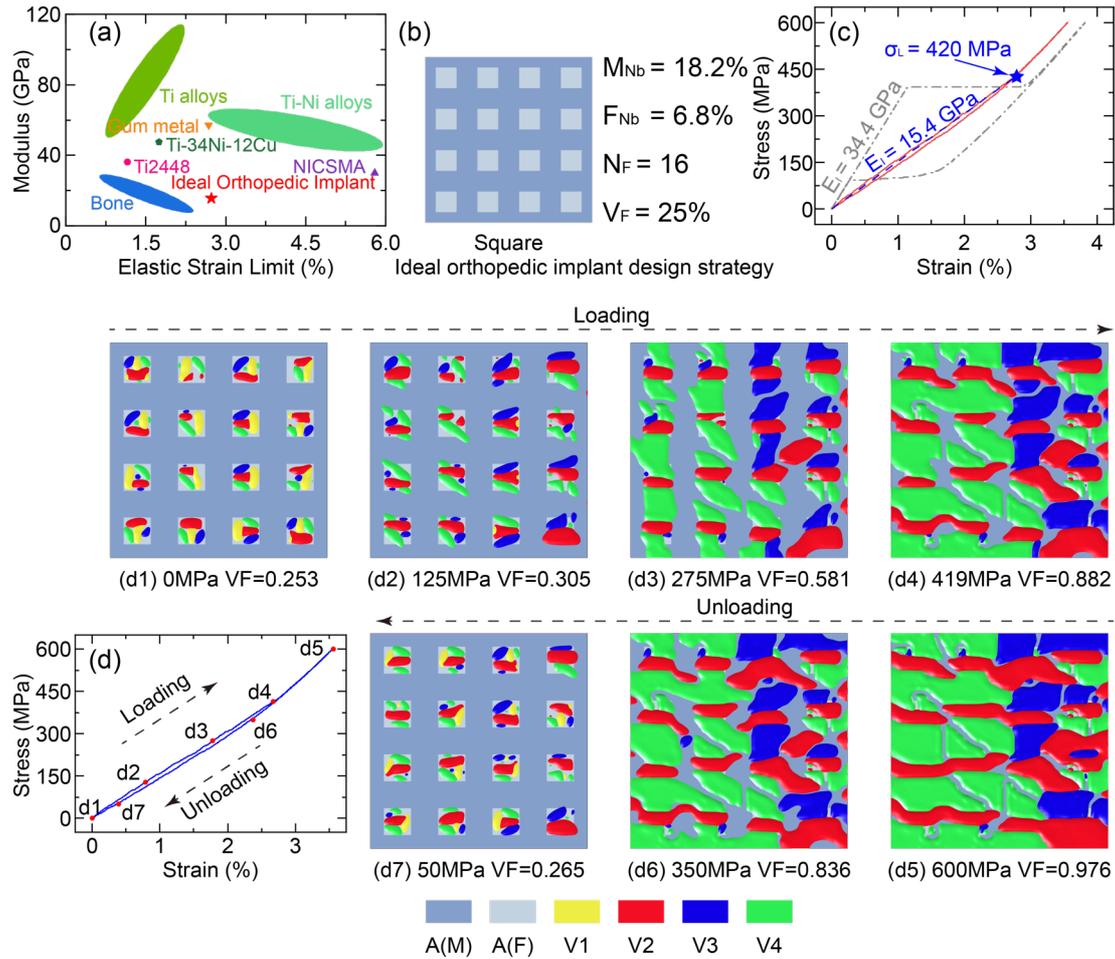

**Fig. 5.** (a) Comparison of the modulus and elastic strain limit between prediction from this study with those reported in the literature. (b) The microstructure of the ideal orthopedic implant obtained from HTP phase-field simulations. (c) SS curves of ideal orthopedic implant (red curve) and Ti2448 (gray curve). (d) Microstructure evolution during MT in ideal orthopedic implant during loading ((d2)–(d4), and unloading ((d5)–(d7)) process.

To gain a thorough understanding of the exceptional mechanical properties, the microstructure evolutions corresponding to the marked points of the loading-unloading SS curve are analyzed, and the results are displayed in Fig. 5d. As shown Fig. 5d1, abundant martensitic particles are observed in the nanofillers, while the matrix still presents an austenite phase when the nanocomposite is quenched to 300 K in a stress-free configuration. These observations are because the martensitic phase is more stable in the Nb-lean fillers ($F_{Nb}$ = 6.8%), as illustrated in Fig. 3a1. The emerged temperature-induced martensitic particles are composed of multi-variants in self-accommodating domain patterns. These martensitic particles behave as seeding beds of martensite, eliminating the nucleation barriers for MT. Upon loading, the stress-induced martensite re-orientation occurs, and the favored variants (such that V2, V3, and V4) dominate and gradually fill the fillers. These martensitic variants gradually and continuously grow into the matrix, which is significantly different from the common avalanche-like discontinuous MT. With increasing stress (Fig. 5d3 and 5d4), the existing martensitic variants continue to expand in matrix, which is also accompanied by the formation and merging of new martensitic configurations in the matrix. At the end of the loading (600 MPa), the martensitic domains spread almost over the entire area of the nanocomposite with a volume fraction of 97.6%, as illustrated in Fig. 5d5, in which the remaining parent phase mainly originates from the martensitic variant interfaces. In the process of unloading, these remaining parent phases initiates the inverse martensite to austenitic transition in the matrix, and the austenite gradually spread from the matrix towards the interfaces of the nanocomposite (Fig. 5d6). The martensite completely disappears in the matrix when the applied stress is reduced to 50 MPa (Fig. 5d7). In addition, the Nb-lean fillers also transform back to the initial self-accommodating multi-variants martensite configurations after unloading, which contributes to the zero-residual strain in a loading-unloading cycle. Therefore, the nanocomposite experiences macroscopically continuous MT throughout the loading and unloading procedure rather than a sharp first-order transition as that in typical SMAs, which thus exhibits linear super-elasticity and ultralow $E_I$ (Fig. 5d). The continuous characteristics of the forward and backward MTs originates from the embedded nanofillers with local heterogeneity that induce non-uniform local stress field associated with the geometrical structure and the variation of MT critical stress due to the change of $C_{Nb}$ as discussed above.

These results thus suggest a new route for achieving desirable mechanical properties to meet the requirement of different applications by nanocomposite engineering. In particular, the designed Ti-Nb nanocomposite realizes a perfect combination of ultra-low modulus, linear super-elasticity, and near-hysteresis-free, which cannot be obtained in common materials and will enable many new advanced applications. For instance, the achieved ultralow $E_I$ in the current Ti-Nb-based alloys without compromising their good biocompatibility and superior corrosion resistance is of particular importance for biomedical applications. The mismatch in the Young's modulus between the common implant materials (~110 GPa) and human bone usually causes stress shielding, leading to bone degradation and implant loosening originated from inhomogeneous stress distribution between the implant and the adjacent bone. The elastic modulus of 15.4 GPa in the designed nanocomposite match closely to that of human bones and thus is promising in overcoming this stress-shielding issue completely. To verify the functional stability of the designed nanocomposite, the repeated loading-unloading cycles are performed for the ideal orthopedic implant. The second SS curve and the corresponding evolution of microstructures are illustrated in Fig S1 in the Supporting Information, which coincide with those in the first one, indicating the mechanical stability of the designed ideal orthopedic implant. In addition, we investigate the influence of the distribution of the nanofillers on the mechanical behavior by performing calculations of triangular and honeycomb-array nanocomposites with the same constituent phases as that of ideal orthopedic implant. As the SS curves and the corresponding microstructures and their evolutions shown in in Fig S2 in the Supporting Information, the mechanical behavior of the nanocomposite is also insensitive to the nanofiller distribution.

## 4. Conclusions

In conclusion, we have developed a highly efficient phase-field simulation framework to accelerate the design of microstructures with desired mechanical properties. HTP phase-field simulations are performed for various Ti-Nb nanocomposites designed by four structural feature variables and it is found that the mechanical responses of the nanocomposites are widely tunable and variable. Based on the combination of HTP phase-field simulations and ML techniques, we design a Ti-Nb alloy nanocomposite with ultralow modulus, linear super-elasticity, and nearly-free hysteresis, which is promising for biomaterial applications. These superior mechanical properties attribute to the non-uniform stress distribution and the modulation of the critical stress that facilitates continuous MT. The proposed strategy by nanostructure engineering to improve the mechanical properties is not limited to improve the mechanical properties of ferroelastics, but is also applicable for the optimization of other functional properties in ferroelectric and ferromagnetic transformations for their strong similarity.

## ACKNOWLEDGMENTS

The authors acknowledge the financial support for Tao Xu from the National Natural Science Foundation of China (Grant No. 11802169).

\*   E-mail: xutao6313@shu.edu.cn
\*\* E-mail: zhangty@shu.edu.cn